\documentclass[aps,prb,reprint,groupedaddress,showpacs]{revtex4-1}
\usepackage{amsmath}
\usepackage{graphicx}
\usepackage{natbib}
\usepackage{bm}
\usepackage{color}
\usepackage{epstopdf}

\begin{document}

\title{Laser fabrication of crystalline silicon nanoresonators from an amorphous film for low-loss all-dielectric nanophotonics}

\author{P.~A.~Dmitriev,$^1$ S.~V.~Makarov,$^{1}$ V.~A.~Milichko,$^{1}$
I.~S.~Mukhin,$^{1,2}$ A.~S.~Gudovskikh,$^{2}$ A.~A.~Sitnikova,$^{3}$ A.~K.~Samusev,$^{1,3}$ A.~E.~Krasnok,$^1$ and P.~A.~Belov$^1$}
\address{
$^{1}$~ITMO University, St.~Petersburg 197101, Russia\\
$^{2}$~St.~Petersburg Academic University, St.~Petersburg 194021, Russia\\
$^{3}$~Ioffe Physical-Technical Institute of the Russian Academy of Sciences, St.~Petersburg 194021, Russia}

\begin{abstract}
The concept of high refractive index subwavelength dielectric nanoresonators, supporting electric and magnetic optical resonances, is a promising platform for waveguiding, sensing, and nonlinear nanophotonic devices. However, high concentration of defects in the nanoresonators diminishes their resonant properties, which are crucially dependent on their internal losses. Therefore, it seems to be inevitable to use initially crystalline materials for fabrication of the nanoresonators. Here, we show that the fabrication of crystalline (low-loss) resonant silicon nanoparticles by femtosecond laser ablation of amorphous (high-loss) silicon thin films is possible. We apply two conceptually different approaches: recently proposed laser-induced transfer and a novel laser writing technique for large-scale fabrication of the crystalline nanoparticles. The crystallinity of the fabricated nanoparticles is proven by Raman spectroscopy and electron transmission microscopy, whereas optical resonant properties of the nanoparticles are studied using dark-field optical spectroscopy and full-wave electromagnetic simulations.
\end{abstract}

\maketitle

\section{Introduction}

High refractive index dielectric nanoparticles with low dissipative losses provide excitation of strong magnetic optical resonances~\cite{Cummer_08, Zhao09, evlyukhin2010, garcia2011strong, KrasnokOE, kuznetsov2012, Brener_12, Lukyanchuk13, krasnok2015towards} and represent a good alternative to high-loss in visible range plasmonic nanoparticles for such applications as electromagnetic field enhancement and sensing~\cite{albella2013low, zambrana2015purcell, samusev2015NL, caldarola2015non}, antireflective coatings~\cite{spinelli2012broadband}, perfect reflectors~\cite{evlyukhin2010, Valentine2014}, light wavefront manipulation~\cite{Staude_15, Kuznetsov2015meta}, superdirective scattering~\cite{KrasnokNanoscale, KrasnokAPLSup2014}, enhancement of nonlinear effects~\cite{ShcherbakovNL2014, makarov2015tuning}. One of the most popular materials for all-dielectric nanophotonics is crystalline silicon (c-Si), because it is a low-cost CMOS compatible semiconductor with large real part and low imaginary part of its refractive index, providing desirable optical response. Indeed, amorphous silicon has much larger imaginary part of its refractive index (up to two orders of magnitude in the visible range)~\cite{palik} compared to the crystalline one.

The most controllable method of silicon film nanopatterning -- reactive ion etching combined with electron-beam lithography -- is limited by microscale size of fabricated structures and their high cost. Therefore, various methods of the resonant silicon nanoparticles fabrication have been developed during last years, including colloidal chemistry~\cite{ShiAdvMat2012}, thin film dewetting~\cite{abbarchi2014wafer} and laser ablation~\cite{kuznetsov2012, chichkov2014NatCom}, where initially crystalline samples or additional annealing were used to achieve crystalline state of the nanoparticles.
\begin{figure}[!b]
\includegraphics[width=0.5\textwidth]{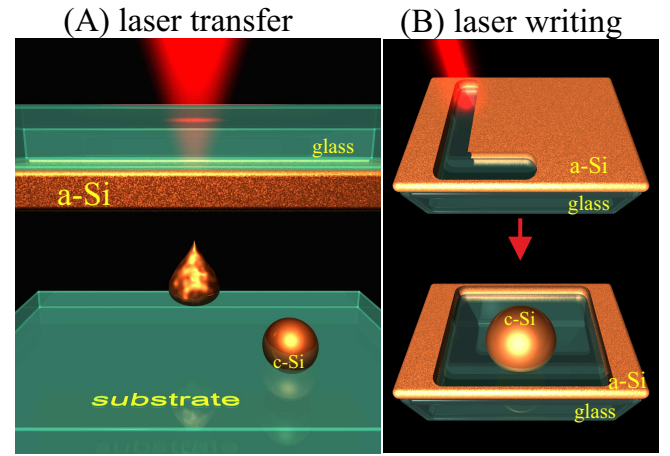}
\caption{Schematic illustration of (A) laser-induced forward transfer and (B) writing of crystalline nanoparticles.}
\label{Fig1}
\end{figure}
\begin{figure*}
\centering
\includegraphics[width=0.99\textwidth]{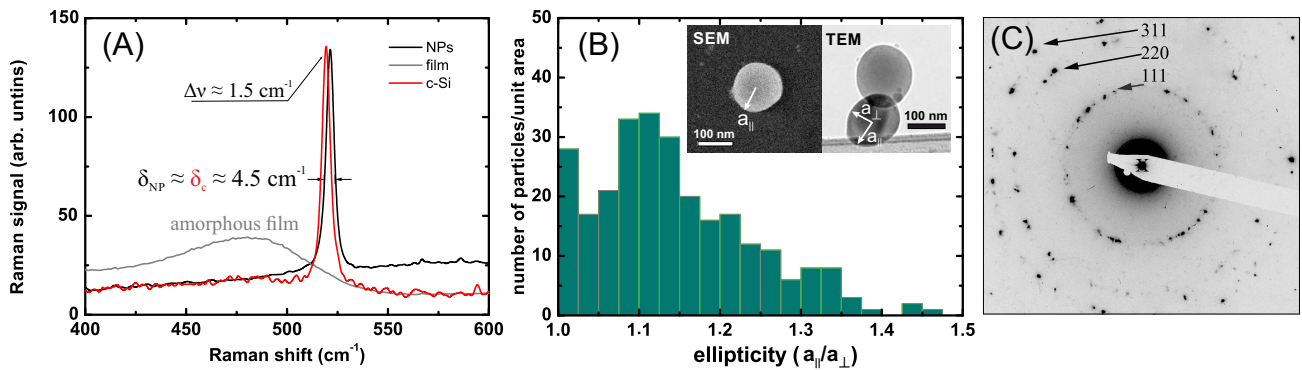}
\caption{(A) Experimental Raman spectra from an individual nanoparticle with a diameter close to 170~nm (black), initial \textit{a}-Si:H film (grey), and reference bulk crystalline silicon (red). (B) Ellipticity distribution of transferred silicon nanoparticles; insets: SEM and TEM images of typical silicon nanoparticles. (C) Electron diffraction picture from the crystalline nanoparticles.}\label{Cryst}
\end{figure*}
However, simple large-scale annealing~\cite{koster1978crystallization} of an amorphous silicon film or nanoparticles is not applicable when a supporting substrate has low temperature of damage (conventional glasses, polymers, etc.) or contains temperature-sensitive nanoobjects (metamaterials, nanowaveguides, etc.). In that cases, pulsed laser annealing has been used for nanoscale crystallisation of amorphous silicon films~\cite{chimmalgi2005nanoscale} and amorphous nanoparticles~\cite{chichkov2014NatCom}.

In this paper, we demonstrate, for the first time to the best of our knowledge, single-step transformation of an amorphous hydrogenated silicon film into crystalline nanoparticles with the magnetic and electric Mie-type resonances by means of two different methods involving laser ablation. The first method is the development of laser-induced transfer technique, the so-called laser-induced forward transfer (LIFT) of nanoparticles~\cite{chichkov2014NatCom} (Fig.~\ref{Fig1}A), and the second one is an original technique, where each nanoparticle is produced by its direct cutting from a continuous amorphous film (Fig.~\ref{Fig1}B). The main advantages of these two methods over those reported previously, that they allow one-stage non-lithographic fabrication of \textit{crystalline} optical nanoresonators on both transparent, and opaque materials.

\section{Results and discussion}

\subsection{Laser-induced forward transfer of c-Si nanoparticles}

Silicon nanoparticles generation and LIFT was carried out by means of a commercial femtosecond laser system (Femtosecond Oscillator TiF--100F, Avesta Poject), emitting laser pulses at a central wavelength of $\lambda\approx$ 800~nm and a pulse duration of 100~fs at a repetition rate of 80~MHz. Selected by a Pockels cell-based pulse picker (Avesta Poject), the laser pulses were tightly focused by an oil immersion microscope objective (Olympus 100$\times$) with a numerical aperture (NA) of 1.4. According to the relation \emph{d}~$\approx1.22\lambda$/NA, an estimated diameter of the beam focal spot is d~$\approx0.7~\mu$m, which is close to d$_{1/e}$~$\approx$~0.68~$\mu$m measured by the standard method~\cite{liu1982simple}.

An 80-nm \textit{a}-Si:H film (initial hydrogen concentration $\sim$10$\%$), deposited on a substrate of fused silica by plasma enhanced chemical vapor deposition from SiH$_{3}$ precursor gas was used as a sample for nanoparticles fabrication. The samples were placed on a three-dimensional air-bearing translating stage driven by brushless servomotors (ABL1000, Aerotech), allowing translation of the sample with an accuracy better than 100~nm.

In all experiments, the nanoparticles were fabricated from a smooth surface (in a single-shot regime) in the forward-transfer (LIFT) geometry (Fig.~\ref{Fig1}A), when the receiving substrate is placed under the film with a spacing of $\sim$50$~\mu$m. This geometry has an advantage over the back-transfer geometry owing to the possibility of nanoparticle transfer onto a wide variety of substrates, including opaque and structured samples. The silicon nanoparticles were fabricated at laser energies E~$<$~2~nJ, providing a laser fluence~(F) range of F = $4 \rm E/\pi d^{2}_{1/e}~<$~550~mJ/cm$^{2}$. The nanoparticles are almost spherical shape and their diameters lie in the range of 50--200~nm, depending on the fluence.

In order to characterize the crystalline structure of the initial film and the produced individual nanoparticles, we provided Raman scattering measurements. The Raman spectra were recorded by a micro-Raman apparatus (AIST-NT SmartSPM system with a Raman spectrometer HORIBA LabRam HR) under excitation by a 632.8-nm HeNe laser through a 100$\times$ microscope objective (NA=0.9) and projected onto a thermoelectrically cooled charge-coupled device (Andor DU 420A-OE 325) with a 600-g/mm diffraction grating. The individual nanoparticles were placed by a piezo stage (AIST-NT) in the center of the focused laser beam (1.8~$\mu$m in diameter) with an accuracy of $\sim$100~nm.

\begin{figure*}
\centering
\includegraphics[width=0.99\textwidth]{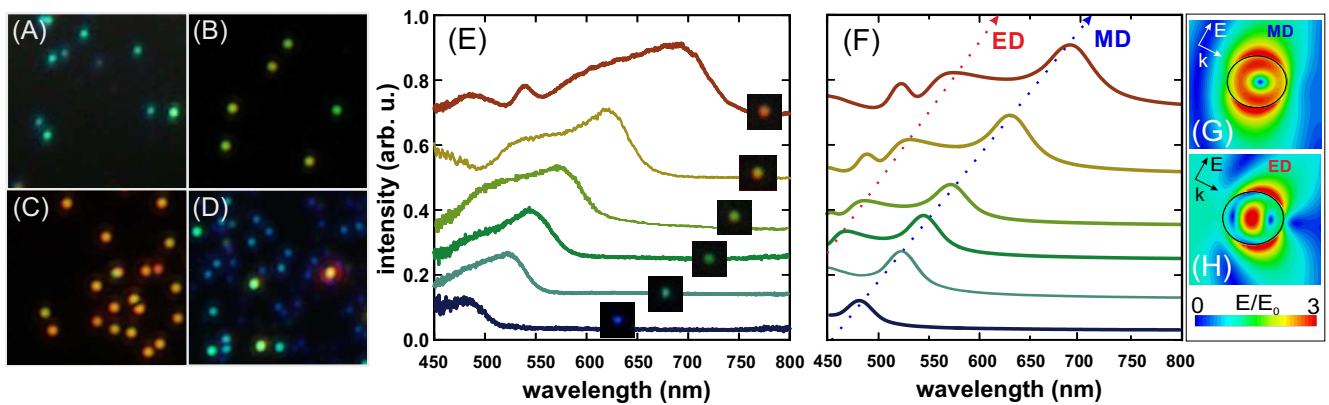}
\caption{Dark-field optical images of silicon nanoparticles fabricated at different peak fluences: 120 (A), 130 (B), 140 (C), 160~mJ/cm$^{2}$ (D). Experimental (E) and theoretical (F) spectra for scattered p-polarized incident light (angle of incidence is 65$^{\circ}$) from individual nanoparticles with the radius parallel to substrate surface a$_{\parallel}$ = 55~nm (blue), 65~nm (spring green), 68~nm (green), 72~nm (olive), 85~nm (yellow) and 92~nm (red) with the ellipticity coefficient of 1.12. Numerically calculated electric field distributions in the silicon nanoparticle with a$_{\perp}$ = 85~nm at the wavelengths of 635~nm (G) and 525~nm (H).}\label{DF}
\end{figure*}

We characterized the initial \textit{a}-Si:H film revealing its amorphous state from the observation of its broad Raman peak centered around 480~cm$^{-1}$. The measured Raman spectra from individual nanoparticles have narrow peaks at 521.5~cm$^{-1}$, corresponding to the crystalline cubic diamond structure. The reference Raman signal from a bulk crystalline silicon wafer and the literature data say that the Raman peak of pure crystal corresponds to 520~cm$^{-1}$. The slight positive shift of the peak of the nanoparticles $\Delta$$\nu$ = 1.5~cm$^{-1}$ is explained by the residual compressive stress~\cite{de1996micro}. Another important characteristic extracted from the Raman spectra is the crystallite size, which is larger than $\sim$20~nm, because the Raman peaks of the nanoparticles have almost the same halfwidth (4--5~cm$^{-1}$) as the peak from bulk crystalline silicon wafer (4.5~cm$^{-1}$)~\cite{campbell1986effects}.

The Raman measurements agree with characterization of the fabricated nanoparticles by means of transmission electron microscopy (TEM). We used specimen grids (3-mm-diameter, 200-mesh copper grids, coated on one side with a 20-nm-thick film of amorphous carbon) to collect nanoparticles ablated from the \textit{a}-Si:H film. The size, structure, and composition of the collected nanoparticles were determined using bright and dark field TEM imaging, see the inset in Fig.~\ref{Cryst}B. The analysis of the electron diffraction pattern from several nanoparticles shows clear maxima, corresponding to certain crystalline planes (Fig.~\ref{Cryst}C). TEM imaging also provides information on the oblateness of the particles along the direction perpendicular to the substrate surface, giving the average ellipticity about $a_{\parallel}/a_{\perp}\approx$1.12, where $a_{\parallel}$($a_{\perp}$) is the particle semi-major (semi-minor) axis oriented parallel (perpendicular) to the surface of the substrate. Indeed, scanning electron microscopy (SEM, Carl Zeiss, Neon 40) confirms that the particles possess axial symmetry along the substrate normal (Fig.~\ref{Cryst}B). These results correlate with previously observed oblateness of transferred silicon nanoparticles~\cite{Lukyanchuk13}.

As was reported previously, the number of nanoparticles~\cite{zywietz2014generation} and their size~\cite{chichkov2014NatCom} strongly depend on laser fluence. In our experiments, we also observed similar behaviour. In particular, two different regimes of the nanoparticle generation were observed. The first regime represents the gradual growth of the nanoparticle size with an increase of fluence up to 150~mJ/cm$^{2}$, manifesting in the change of their colors from blue to red (Figs.~\ref{DF}A--C). At F~$\approx$~120, 130, and 140~mJ/cm$^{2}$ the mean sizes are about 130$\pm$10, 150$\pm$15, and 180$\pm$15~nm, respectively; whereas the average number of nanoparticles are about 2.5$\pm$0.5, 2.7$\pm$0.8, and 3.1$\pm$1.0, respectively. Such dependencies can be described in terms of the spallation mechanism of laser ablation, where a thin molten layer is spalled due to the laser-induced tensile pressure waves~\cite{ionin2013thermal,wu2014microscopic}, breaking into a number of liquid droplets via the Rayleigh-Plateau instability~\cite{papageorgiou1995breakup}. The photomechanically spalled volume increases as V$\sim$ln(E) under the action of a Gaussian beam, owing to the logarithmic dependence of the spalled surface layer area r$^{2}$$_{s}$$\sim$ln(E)~\cite{bauerle2013laser}, whereas the maximum thickness of the layer remains almost constant~\cite{ionin2013thermal}. The increase of the molten volume led to an increase in the number of nanoparticles~\cite{zywietz2014generation} and their size~\cite{chichkov2014NatCom}, which agrees with our results (Fig.~\ref{DF}A--C).

The second regime of the nanoparticle fabrication corresponds to a fluences of F~$>$~150~mJ/cm$^{2}$, where large (red) nanoparticles are accompanied by small nanoparticles with a much broader size distribution (Fig.~\ref{DF}D). This regime is related to unstable boiling of superheated silicon~\cite{bulgakova2001pulsed, ionin2013thermal, wu2014microscopic}, when the nanoparticle formation occurs through explosive decomposition of the material into vapor and small clusters/droplets~\cite{itina2009molecular,wu2014microscopic}, yielding nanoparticles with the mean sizes smaller than 100~nm~\cite{amoruso2004generation}. Small silicon nanoparticles fabrication in this regime has been extensively studied over last two decades~\cite{amoruso2004generation, tull2006formation} for a number of biomedical applications. However, one can conclude that the high-fluences regime, yielding very broad size distribution (dispersion is $>$ 100\%, see Fig.~\ref{DF}D),
is not desirable for reproducible Mie-type nanoresonators fabrication, whereas the low-fluence regimes (Fig.~\ref{DF}A--C) provide much more controllable nanoparticles transfer. The Raman, TEM and SEM characterization of the tranferred silicon nanoparticles helped to determine their almost perfect crystalline phase, meaning that it is possible to fabricate the crystalline silicon nanoresonators from amorphous films, which is very useful for low-loss all-dielectric nanophotonics.
\begin{figure*}
\centering
\includegraphics[width=0.99\textwidth]{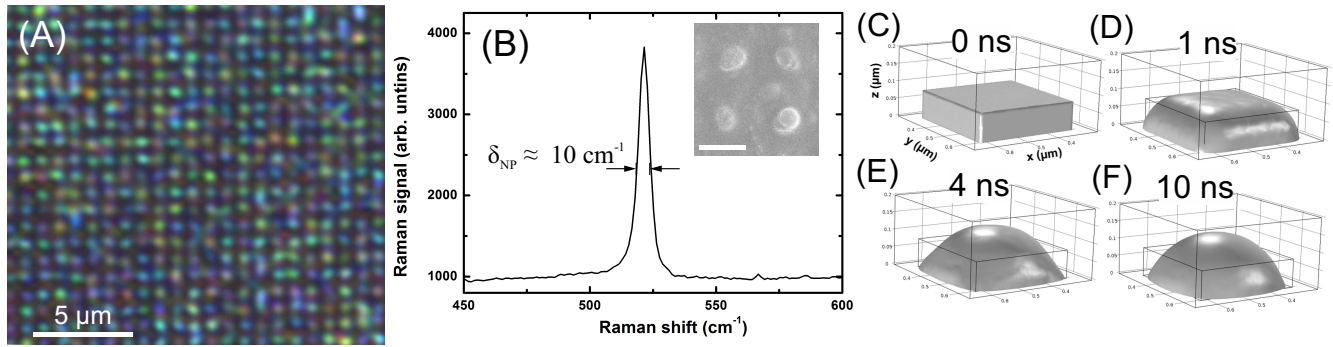}
\caption{(A) Optical image of an array of the silicon nanoparticles fabricated by direct laser writing at the fluence of 100~mJ/cm$^{2}$, repetition rate of 80~MHz and scanning period of 0.9~$\mu$m. (B) Raman spectrum of a silicon nanoparticle from the array; the inset displays SEM image of the written nanoparticles covered by a 10-nm gold layer with the scale bar of 700~nm. Numerically modelled evolution of a liquid silicon patch with a height of 80~nm and equal sides of 300~nm on a solid fused silica substrate at the different time-steps: 0~ns (C), 1~ns (D), 4~ns (E), and 10~ns (F).}\label{Writting}
\end{figure*}

In order to demonstrate the excellent resonant properties of the particles, we provide polarization-resolved scattering spectra measurements in a home-made dark-field scheme, with the nanoparticles illuminated by p-polarized light from a halogen lamp (HL--2000--FHSA) at the angle of 65$^{\circ}$ to the surface normal, and scattered signal collection was performed by means of a Mitutoyo M Plan APO NIR 50x objective (NA=0.45). Changing of the incident light polarization does not affect the resonances positions in the acquired spectra, which is discussed elsewhere~\cite{permyakov2015probing}. The scattered light was registered by the same spectrometer as in the Raman measurements with a 150-g/mm diffraction grating.

The excitation of the electric dipole (ED) and magnetic dipole (MD) resonances is proven by full-wave electromagnetic simulations in CST Microwave Studio. The scattering geometry is modelled as a c-Si ellipsoid with two different axis ($a_{\perp}$ and $a_{||}$) and the fixed ellipticity $a_{\parallel}/a_{\perp}\approx$~1.12, i.e. for the most probable ellipticity parameter in the experiments. The ellipsoid is irradiated by a plane wave at the angle 65$^{\circ}$ in vacuum. Modeling of scattering spectra (Fig.~\ref{DF}F) shows good agreement with the corresponding experimental ones (Fig.~\ref{DF}E), whereas the modeled electric field distributions in the ellipsoids at different wavelengths reveal excitation of MD (Fig.~\ref{DF}G) and ED (Fig.~\ref{DF}H). Some deviations between the shapes of theoretical and experimental spectra arise from the glass substrate and presence of a thin natural oxide layer on the nanoparticles~\cite{evlyukhin2012demonstration, ginzburg14}. Such optical and TEM/SEM analysis of the crystallinity and the average ellipticity of the nanoparticles will be useful for further applications of the LIFT method in all-dielectric nanophotonics.

\subsection{Laser writing of c-Si nanoparticles}

As we have shown, the LIFT method allows to fabricate c-Si nanoparticles with distinguished resonances, however it does not allow to obtain an ordered array of nanoparticles from the amorphous film similarly to the previously published results on bulk c-Si~\cite{zywietz2014generation}. Therefore, we develop a novel method of direct laser writing of c-Si resonant nanoparticles via cutting of submicron square patches by the fs-laser irradiation at a 80~MHz repetition rate (Fig.~\ref{Fig1}A). At a scanning velocity of 1~mm/s and F~$\approx$~100~mJ/cm$^{2}$, narrow (a width of $\approx$~300~nm) grooves can be written directly in the \textit{a}-Si:H film.

Under the optimal conditions of fabrication, the resulting array of nanoparticles has a period of about 0.9~$\mu$m. Bright colors in Fig.~\ref{Writting}A originate from the resonant scattering of each nanoparticle similarly to the individual nanoparticles made by LIFT (Fig.~\ref{DF}). Since the written arrays are quite dense and have some size dispersion ($\approx$~30~\%),
the comprehensive experimental analysis of individual nanoparticles optical properties is difficult. However, their images look similarly to the nanoparticles produced by means of the LIFT method (Figs.~\ref{DF}A--E). Though the cutting should produce an array of square patches, our SEM images show that the nanoparticles look almost circular from the top (Fig.~\ref{Writting}B). This is due to thermal isolation of the patch from the rest of the film and its overheating during the cutting by a train of femtosecond laser pulses with a period of 12.5~ns. These microscale patches are known to be unstable at high temperatures and undergo dewetting to a certain number of similar nanoparticles depending on their dimensions~\cite{thompson2012solid}. In order to provide deeper insight into the mechanism of the patches reshaping, we model the time dynamics of the cut liquid silicon patch with a height of 80~nm and similar widths of 300~nm (Figs.~\ref{Writting}C) on a fused silica substrate in the COMSOL software, solving the incompressible Navier-Stokes equations and taking into account the parameters of the used materials. The modeling shows that after ten nanoseconds the patch is transformed into the hemisphere with a height of about 140~nm and width of about 350~nm (Figs.~\ref{Writting}C-F), giving qualitative agreement with the experimentally observed shapes.

The corresponding Raman signals from these nanoparticles also reveal the crystalline state of the nanoparticles written by the laser, demonstrating a narrow peak at 520~cm$^{-1}$, with a halfwidth about 10~cm$^{-1}$ (Fig.~\ref{Writting}B), which is larger than the halfwidth (about 4--5~cm$^{-1}$) of the nanoparticles fabricated by the LIFT method. The larger halfwidth corresponds to the smaller mean crystallite size, i.e. less than 10~nm according to the previous studies~\cite{campbell1986effects}. The origin of this difference is related to the faster cooling rate for the written nanoparticles on the fused silica substrate as compared with nanoparticles flying in air before deposition in the LIFT mode, owing to the 20-fold larger value of the glass thermal conductivity in comparison with that one for air.

\section{Conclusions}

In summary, we have experimentally proven that the laser fabrication of crystalline (low-loss) silicon nanoparticles, supporting magnetic Mie-type resonance, can be performed from amorphous (high-loss) \textit{a}-Si:H films, making it unnecessary to use initially crystalline material or additional annealing. The fabrication has been carried out in two modes: laser-induced forward transfer (LIFT) and laser writing of the nanoparticles. The comparative analysis of these two methods shows that the laser-induced transfer produces crystalline nanoparticles with similar dimensions at certain fluences, whereas laser writing yields polycrystalline nanoparticles, which are ordered in a periodic array. The developed LIFT method enables controllable deposition of \textit{crystalline} silicon nanoparticles onto \textit{virtually any substrate} (metal, semiconductor, metamaterial, nanostructured etc.), unlike previous laser approaches for high-index nanoparticles.

\section*{Acknowledgments}

This work was financially supported by Russian Science Foundation (Grant~15-19-00172). We are also thankful to Sergey Kudryashov, Boris Chichkov and Arseniy Kuznetsov and Dmitriy Zuev for useful comments and discussions.


%

\end{document}